\documentclass[a4paper,12pt]{article}
\usepackage[section] {placeins}
\usepackage{morefloats}
\usepackage{achemso}
\usepackage{epstopdf}
\usepackage{bbm}
\usepackage{textcomp}
\usepackage{amsfonts}
\usepackage{amssymb}
\usepackage{amsmath}
\usepackage{pgf}
\usepackage{setspace}
\usepackage[top=3cm,left=1.5cm,right=1.5cm,bottom=2cm,foot=1cm]{geometry}

\usepackage{multirow}

\usepackage{isomath}
\usepackage{mathtools}
\newcommand{\mvec}[1]{\vectorsym{#1}}

\usepackage[normalem]{ulem}

\newcommand{\RNum}[1]{\uppercase\expandafter{\romannumeral #1\relax}}

\begin{document}
\title{Energy Decomposition along Reaction Coordinate and Energy Weighted Reactive Flux: \RNum{1}. Theory}
\author{Wenjin Li\\
Institute for Advanced Study, Shenzhen University, Shenzhen, China\\
liwenjin@szu.edu.cn}
\date{}
\maketitle

\begin{abstract}
A theoretical framework is proposed for an energy decomposition scheme along the reaction coordinate, in which the ensemble average of the potential energy weighted with reactive flux intensity is decomposed into energy components at the per-coordinate level. The decomposed energy quantity is demonstrated to be closely related to the free energy along the reaction coordinate and its connection to the emergent potential energy is provided. We also explore the property of the reactive flux weighted by the potential energy in the subspace of collective variables. A free energy analogue is then proposed in the subspace of collective variables and the directional derivative of this free energy analogue in the direction of the reactive flux is shown to be useful in the identification of the reaction coordinate and the minimum free energy path.      
\end{abstract}

\section*{\RNum{1}. INTRODUCTION}
Complex biomolecular systems, such as protein folding, large conformational changes, and protein-ligand interactions, have been intensively investigated by molecular dynamics (MD) simulations in the past several decades~\cite{karplus2002molecular, wang2021molecular}. In spite of the great success in unveiling the detailed mechanisms of biomolecular systems, biomolecular simulations face mainly the challenges in several aspects such as the force field, the sampling problem, and the extraction of insightful information from high-dimensional trajectories. In the endeavour of overcoming the latter two challenges, a complex system from a physical viewpoint is usually reduced to a prototypical system consisting of two metastable states that are separated by a high energy barrier~\cite{Chandler:1978aa}. In a straight-forward simulation, the system will spend most of the time in the metastable states and the transitions from one metastable state to the other are rarely sampled. To tackle the sampling issue, one common solution among many others resorts to a large ensemble of short trajectories obtained via various path sampling strategies~\cite{pan2008finding, Allen:2009aa,Borrero:2007aa,faradjian2004computing, elber2017new, berezhkovskii2019committors,zuckerman2017weighted,Dellago:1998aa, Bolhuis:1998aa, Dellago:1999aa, Bolhuis:2002aa, Dellago:2002aa}. These strategies usually possess the following properties: (1) Avoid redundant sampling of metastable basins and enhance the sampling of transition or barrier regions; (2) Easy to be parallelized and thus best utilize supercomputers; (3) The sampled trajectories are natural and form a non-equilibrium ensemble (no bias potential is introduced but the weight of a trajectory is different to the one in an equilibrium ensemble). Examples of such an ensemble of trajectories are swarms of trajectories~\cite{pan2008finding,roux2021string} and transition path ensemble (TPE)~\cite{Dellago:1998aa, Bolhuis:1998aa, Dellago:1999aa, Bolhuis:2002aa, Dellago:2002aa}. 

Since a non-equilibrium ensemble of short trajectories is generally more affordable for biomolecular systems than a long equilibrium ensemble, it is of great interest to extract from it useful kinetic and dynamic information, which includes for example (1) the metastable states and the matrix of transition probabilities among these metastable states; (2) the reaction coordinate and the free energy along it; (3) the contributions of different parts of the biomolecule (a group of atoms or residues) to the rare transition. Here, we will pay a special attention to the latter two cases. For the identification of reaction coordinate, the committor function of a configuration, that is the probability of the trajectories initiated from the configuration will reach the product state first before visiting the starting state, is generally viewed as the `ideal' reaction coordinate~\cite{li2014recent}. For a diffusive process with quadratic energy barrier, the gradient of the committor was demonstrated to be parallel to the eigenvector of the only negative eigenvalue of the matrix $\mvec{HD}$~\cite{berezhkovskii2005one, rhee2005one}, where $\mvec{H}$ is the Hessian matrix of the potential energy and $\mvec{D}$ is the diffusion tensor. Using the committor as the reference, several approaches such as genetic neural network method~\cite{Ma:2005aa} and likelihood maximization method~\cite{Peters:2006aa, Peters:2007aa, Peters:2013aa} have been proposed to select the reaction coordinate out of a pool of candidate coordinates. However, the evaluation of the committor is computationally expensive, although an efficient method to estimate the committor via a fitting procedure is available~\cite{li2015reducing}, approaches were then developed to extract the reaction coordinate out of a non-equilibrium ensemble of trajectories without extra simulations~\cite{Li:2016aa, li2018equipartition, li2021optimizing}. Li and co-workers suggested that the emergent potential energy along a coordinate is useful in appraising the relevance of the coordinate to the reaction coordinate~\cite{Li:2016aa}. Later on, the equipartition terms in the TPE along a coordinate were maximized to obtain a one-dimensional representation of the reaction coordinate~\cite{li2018equipartition}. Recently, a density-weighted average of the flux along a coordinate is shown to be a promising quantity to select coordinates most relevant to the reaction coordinate~\cite{li2021optimizing}, and the theoretical framework for this quantity and its connection to the transition path theory is established~\cite{li2022time}.

Existing methods that estimate the contributions from different parts of the system in the transition from one metastable state to the other largely rely on free energy decomposition. Typical examples are free energy perturbation~\cite{boresch1994free}, thermodynamic integration~\cite{brady1995decomposition}, and the MM/PB(GB)SA approach~\cite{srinivasan1998continuum, gohlke2004converging}. However, the decomposition of free energy was carried out either along an artificial pathway or in concern of only the two end states. On the other hand, the energy decomposition depends on the very path of integration~\cite{smith1994free}. Thus, the reaction coordinate is considered to be the natural choice of the path along which the energy decomposition should be performed. Following this thinking path, the potential energy was then decomposed along the committor at the single-coordinate level in the TPE~\cite{}. This so-called emergent potential energy approach was extended by Li into a residue-residue mutual work analysis approach~\cite{li2020residue}, in which the energy contribution is quantified by a so-called ensemble averaged work (EAW) and the EWA on a residue is simply the summation of the per-coordinate EAW over all the coordinates of the residue. Importantly, the EWA on a residue due to its interaction with another residue is provided by decomposing the atomic force on an atom into pairwise forces (the force on one atom from its interaction with another atom). The energy contribution on a residue from another one was shown to be different to the one on the latter residue due to the former one. We would like to emphasize that the MM-GBSA method assumes the residue-residue mutual works from each other are the same and the interaction energy between two residues are evenly divided in the per-residue decomposition. In the residue-residue mutual work approach~\cite{li2020residue}, the energy contribution of a residue can be further decomposed into contributions due to translational, rotational, and internal motions of the residue.

Here, inspired by the principal curve defined in the transition path theory~\cite{vanden2010transition}, a theoretical framework for the TPE, we proposed an ensemble average of any quantity weighted with reactive flux or current intensity over a non-equilibrium ensemble in which the reactive current is non-zero. By considering such a weighted ensemble average of the potential energy, an energy decomposition scheme along the reaction coordinate is derived and the properties of the energy components on various coordinates are discussed. Later, we define a reactive flux weighted by the potential energy in the subspace of collective variables (CVs), from which a multi-dimensional free energy analogue of CVs is then derived. In addition, we will show that the directional derivative of this free energy analogue in the direction of the reactive flux  can be used in identifying the reaction coordinate and the minimum free energy path. 

\section*{\RNum{2}. THEORY}
\subsection*{A. Flux and current}
Let $\mvec{x}(t)$ be an infinitely long trajectory of a system that consists of two stable states A and B, where $\mvec{x}$ represents the position coordinates of the system. In this long trajectory, the trajectory pieces that initiate from configurations in the state A and commit to the state B without returning to the state A are known to be reactive trajectories from A to B, which form the so-called TPE. We denote $U$ as an arbitrary ensemble, which could be the TPE or other non-equilibrium ensembles taken from the long trajectory and $\xi(\mvec{x})$ as a generalized coordinate. From a theoretical framework proposed in Ref.\,[\cite{li2022time}] (here the notation and nomenclature in this reference are largely followed), the flux at a fixed location $\xi'$ of $\xi(\mvec{x})$ in the ensemble $U$ is defined as

\begin{equation}
F^{U}(\xi')= \int_{0}^{T}\dot{\xi}(\mvec{x}(t))\delta(\xi(\mvec{x}(t))-\xi')h_{U}(t)dt
\label{eq:flux:xi2}
\end{equation}
where $h_{U}(t)$ is an indicator function for the ensemble $U$, that is $h_{U}(t)=1$ if $\mvec{x}(t)$ belongs to $U$ and $h_{U}(t)=0$ otherwise.

If $\mvec{y}\equiv \{y_{1},y_{2},\cdots,y_{n}\}$ is a set of CVs, which are functions of $\mvec{x}(t)$, the current (from now on we use current when referring to a vector field and flux when a scalar is referred) in the space of CVs in the ensemble $U$ can then be defined as

\begin{equation}
\mvec{J}^{U}(\mvec{y}) = \int_{0}^{T}\dot{\mvec{y}}(t)\delta(\mvec{y}(t)-\mvec{y})h_{U}(t)dt 
\label{eq:currentU:CV}
\end{equation}

\subsection*{B. Ensemble average weighted with current intensity}
Given $f(\mvec{x})$ being an arbitrary function of $\mvec{x}$, we consider a particular ensemble average of the quantity $f(\mvec{x})$ over the surface $\xi(\mvec{x})=\xi'$ in the ensemble $U$, 

\begin{equation}
\langle f(\mvec{x}) \rangle_{\xi(\mvec{x})=\xi',U}^{Flux}\equiv\frac{\int_{0}^{T}f(\mvec{x}(t))\dot{\xi}(\mvec{x}(t))\delta(\xi(\mvec{x}(t))-\xi') h_{U}(t)dt}{\int_{0}^{T}\dot{\xi}(\mvec{x}(t))\delta(\xi(\mvec{x}(t))-\xi') h_{U}(t)dt}
\label{eq:fluxave:f}
\end{equation}
It is inspired by the principal curve proposed in the transition path theory~\cite{vanden2010transition} and is actually the average over the surface $\xi(\mvec{x})=\xi'$ of $f(\mvec{x})$ that is weighted with the current intensity passing through the point $\mvec{x}$ on the surface $\xi(\mvec{x})=\xi'$, that is,

\begin{equation}
\langle f(\mvec{x}) \rangle_{\xi(\mvec{x})=\xi',U}^{Flux}=\frac{\int f(\mvec{x}) \mvec{J}^{U}(\mvec{x})\cdot \nabla \xi(\mvec{x})\delta (\xi(\mvec{x})-\xi')d\mvec{x}}{\int \mvec{J}^{U}(\mvec{x})\cdot \nabla \xi(\mvec{x})\delta (\xi(\mvec{x})-\xi')d\mvec{x}}
\label{eq:fluxave:f2}
\end{equation}

Since a point on the principal curve given in Eq.\,(63) in Ref.\,[~\cite{vanden2010transition}] can be viewed as a conditional canonical expectation of points in an isosurface along the principal curve itself under the assumptions: (1) the committor can be approximated by a function of the principal curve, (2) the curvature effects of the isosurface is negligible, and (3) the overdamped dynamics is assumed. In the following, we will show that $\langle f(\mvec{x}) \rangle_{\xi(\mvec{x})=\xi',U}^{Flux}$ is a conditional canonical expectation of $f(\mvec{x})$ under similar assumptions. For a diffusion process where the Smoluchowski equation is applicable, the reactive current is given by~\cite{johnson2012characterization}

\begin{equation}
\mvec{J}_{R}(\mvec{x})=\rho(\mvec{x})\mvec{D}(\mvec{x})\nabla q(\mvec{x})
\label{eq:flux:OD}
\end{equation}
where $\rho(\mvec{x})$ is the equilibrium probability density in canonical ensemble, $\mvec{D}(\mvec{x})$ is the position-dependent diffusion matrix, and $q(\mvec{x})$ is the committor function. If $\xi(\mvec{x})$ is a monotonically increasing function of the committor, i.e., $q(\mvec{x})=g(\xi(\mvec{x}))$ and thus $\nabla q=g'(\xi)\nabla\xi$, and $U$ is chosen to be the TPE (then $\mvec{J}^{U}(\mvec{x})$ is proportional to $\mvec{J}_{R}(\mvec{x})$), we have,

\begin{subequations}
\begin{align}
\langle f(\mvec{x}) \rangle_{\xi(\mvec{x})=\xi',U}^{Flux}=& \frac{\int f(\mvec{x}) \rho(\mvec{x})g'(\xi(\mvec{x}))\nabla \xi(\mvec{x})\cdot \mvec{D}(\mvec{x})\nabla \xi(\mvec{x})\delta (\xi(\mvec{x})-\xi')d\mvec{x}}{\int \rho(\mvec{x})g'(\xi(\mvec{x}))\nabla \xi(\mvec{x})\cdot \mvec{D}(\mvec{x})\nabla \xi(\mvec{x})\delta (\xi(\mvec{x})-\xi')d\mvec{x}} \\
 =& \frac{\int f(\mvec{x}) \rho(\mvec{x})\nabla \xi(\mvec{x})\cdot \mvec{D}(\mvec{x})\nabla \xi(\mvec{x})\delta (\xi(\mvec{x})-\xi')d\mvec{x}}{\int \rho(\mvec{x})\nabla \xi(\mvec{x})\cdot \mvec{D}(\mvec{x})\nabla \xi(\mvec{x})\delta (\xi(\mvec{x})-\xi')d\mvec{x}}
\end{align}
\label{eq:fluxave:f3}
\end{subequations}
Under the assumption of a local transition tube, in which most of the reactive current are located, the curvature of the surface $\xi(\mvec{x})=\xi'$ in the transition tube could be negligible and the diffusion on it may be approximately position independent (not necessarily isotropic), then $\nabla \xi(\mvec{x})\cdot \mvec{D}(\mvec{x})\nabla \xi(\mvec{x})$ is approximately a constant and can be cancelled out in the above fraction. We thus arrive,

\begin{equation}
\langle f(\mvec{x}) \rangle_{\xi(\mvec{x})=\xi',U}^{Flux}\approx \frac{\int f(\mvec{x}) \rho(\mvec{x})\delta (\xi(\mvec{x})-\xi')d\mvec{x}}{\int \rho(\mvec{x})\delta (\xi(\mvec{x})-\xi')d\mvec{x}}
\label{eq:ave:eq}
\end{equation}

Although we assume $f(\mvec{x})$ being a function of $\mvec{x}$, the weighted ensemble average can be generalized to any time series $f(t)$ extracted from trajectories and Eq.\,(\ref{eq:fluxave:f}) is reformulated into

\begin{equation}
\langle f(t) \rangle_{\xi(\mvec{x})=\xi',U}^{Flux}\equiv\frac{\int_{0}^{T}f(t)\dot{\xi}(\mvec{x}(t))\delta(\xi(\mvec{x}(t))-\xi') h_{U}(t)dt}{\int_{0}^{T}\dot{\xi}(\mvec{x}(t))\delta(\xi(\mvec{x}(t))-\xi') h_{U}(t)dt}
\label{eq:fluxave:fn}
\end{equation}

In case that $\xi(\mvec{x})$ is a good one-dimensional representation of the reaction coordinate and the flux along it ($F^{U}(\xi)$) presents a plateau region, in which $F^{U}(\xi)$ is a constant and equals the number of transition paths $N_{\rm T}$. We then have (Appendix A)  

\begin{equation}
\frac{d\langle f(\mvec{x}) \rangle_{\xi(\mvec{x})=\xi',U}^{Flux}}{d\xi'}=\langle \frac{df(\mvec{x}(t))}{dt}\frac{1}{\dot{\xi}(\mvec{x}(t))} \rangle_{\xi(\mvec{x})=\xi',U}^{Flux} \quad \text{for $\xi'$ in the plateau region}
\label{eq:fluxave:df2}
\end{equation}
In case that $\xi(\mvec{x})$ is the committor function or a monotone of the committor, Eq.~(\ref{eq:fluxave:df2}) holds everywhere. We will show the usefulness of Eqs.~(\ref{eq:ave:eq}) and (\ref{eq:fluxave:df2}) in the next subsection. 

\subsection*{C. Energy decomposition along reaction coordinate}
Given an independent and complete set of configurational coordinates $\mvec{z}\equiv\{z_{\rm 1},z_{\rm 2},\cdots ,z_{\rm n}\}$, we consider a special case that $f(\mvec{z})$ is the potential energy $V(\mvec{z})$ and then

\begin{equation}
\langle \frac{dV(\mvec{z}(t))}{dt}\frac{1}{\dot{\xi}(t)} \rangle_{\xi(\mvec{x})=\xi',U}^{Flux}=\sum_{i}\langle \frac{\partial V(\mvec{z}(t))}{\partial z_{\rm i}}\frac{\dot{z}_{\rm i}(t)}{\dot{\xi}(t)} \rangle_{\xi(\mvec{x})=\xi',U}^{Flux}
\label{eq:ener:decomp}
\end{equation}
Here, we avoid the explicit expressions of $\mvec{z}$ as functions of $\mvec{x}$ to simply the notations. Note that the dimension of $\frac{\partial V(\mvec{z}(t))}{\partial z_{\rm i}}\frac{\dot{z}_{\rm i}(t)}{\dot{\xi}(t)}$ is the same as the one of the generalized force corresponding to the generalized coordinate $\xi$, its integral over $\xi$ gives terms of the same dimension as energy, we thus formally defined the energy decomposed onto a coordinate $z_{\rm i}$ along the coordinate $\xi$ as

\begin{equation}
V_{z_{\rm i},\xi}=\int \langle \frac{\partial V(\mvec{z}(t))}{\partial z_{\rm i}}\frac{\dot{z}_{\rm i}(t)}{\dot{\xi}(t)} \rangle_{\xi(\mvec{x})=\xi',U}^{Flux}d\xi'
\label{eq:ener:comp}
\end{equation}

In the following, we will discuss the properties of the energy component in Eq.\,(\ref{eq:ener:comp}) in various cases under the conditions: (1) The ensemble $U$ is chosen to be the TPE; (2) The system can be described as a reaction coordinate in interaction with a large number of bath modes~\cite{zwanzig1973nonlinear}; (3) The behaviours of bath modes in the TPE resemble the ones in the equilibrium ensemble.

\subsubsection*{1. When $\xi$ is the reaction coordinate}
We first consider an ideal case that $\xi(\mvec{x})$ is the reaction coordinate and it belongs to $\mvec{z}$, then the energy component on the reaction coordinate $\xi$ along itself is

\begin{equation}
V_{\xi,\xi}=\int \langle \frac{\partial V(\mvec{z}(t))}{\partial \xi} \rangle_{\xi(\mvec{x})=\xi',U}^{Flux}d\xi'\approx \int \langle \frac{\partial V(\mvec{z})}{\partial \xi} \rangle_{\xi(\mvec{x})=\xi'}d\xi'
\label{eq:comp:RC}
\end{equation}
Here $\langle \cdots \rangle_{\xi(\mvec{x})=\xi'}$ stands for the conditional canonical average over the surface $\xi(\mvec{x})=\xi'$ and Eq.\,(\ref{eq:ave:eq}) is used. $\langle \frac{\partial V(\mvec{z})}{\partial \xi} \rangle_{\xi(\mvec{x})=\xi'}$ is closely related to the potential of mean force $A$ along $\xi$ as~\cite{darve2007therm}

\begin{equation}
\frac{dA(\xi)}{d\xi}|_{\xi(\mvec{x})=\xi'} = \langle \frac{\partial V}{\partial \xi}\rangle_{\xi(\mvec{x})=\xi'} - \langle k_{\rm B}T\frac{\partial \ln|\mvec{J}|}{\partial\xi} \rangle_{\xi(\mvec{x})=\xi'}
\label{eq:free:ener}
\end{equation}
with $\mvec{J}$ being the Jacobian matrix upon transforming from Cartesian to generalized coordinates, $k_{\rm B}$ being the Boltzmann's constant and $T$ the temperature. If the entropic contribution due to the change of volume element associated with the generalized coordinates ($-\langle k_{\rm B}T\frac{\partial \ln|\mvec{J}|}{\partial\xi} \rangle_{\xi(\mvec{x})=\xi'}$) is negligible, the energy component $V_{\xi,\xi}$ will approximates the potential of mean force along $\xi$. 

The energy component on a coordinate other than the reaction coordinate (i.e., a bath mode) along the reaction coordinate $\xi$ is

\begin{equation}
\frac{dV_{z_{\rm i},\xi}}{d\xi}|_{\xi(\mvec{x})=\xi'} = \frac{1}{F^{U}(\xi')}\int_{0}^{T}\frac{\partial V(\mvec{z}(t))}{\partial z_{\rm i}}\dot{z}_{\rm i}(t)\delta(\xi(\mvec{x}(t))-\xi')h_{U}(t)dt \approx 0 \quad \text{for $z_{\rm i}\neq \xi$}  
\label{eq:comp:bath}
\end{equation}
Thus, the energy component on a bath mode along the reaction coordinate vanishes as the distribution of the generalized velocity of a bath mode resembles the equilibrium one.

What if none of the coordinates in $\mvec{z}$ is the true reaction coordinate? Let us assume without loss of generality that the reaction coordinate $\xi$ is a function of $z_{\rm 1}$ and $z_{\rm 2}$ and $\eta$ is the bath mode that is orthogonal to $\xi$ in the subspace of $z_{\rm 1}$ and $z_{\rm 2}$, then the energy component on $z_{\rm j}$ ($j\in \{1,2\}$) along $\xi$ is 

\begin{subequations}
\begin{align}
\frac{dV_{z_{\rm j},\xi}}{d\xi}|_{\xi(\mvec{x})=\xi'} &= \frac{1}{F^{U}(\xi')}\int_{0}^{T}\frac{\partial V(\mvec{z}(t))}{\partial z_{\rm j}}[\frac{\partial z_{\rm j}}{\partial \xi}\dot{\xi}(t)+\frac{\partial z_{\rm j}}{\partial \eta}\dot{\eta}(t)]\delta(\xi(\mvec{x}(t))-\xi')h_{U}(t)dt \\
&\approx \langle \frac{\partial V(\mvec{z}(t))}{\partial z_{\rm j}}\frac{\partial z_{\rm j}}{\partial \xi} \rangle_{\xi(\mvec{x})=\xi',U}^{Flux}
\end{align}
\label{eq:comp:mix}
\end{subequations}
Again, the energy term related to the bath mode $\eta$ vanishes by assumption. Thus, the energy component on a mixed coordinate (a coordinate that is a combination of the reaction coordinate and a bath mode) along the reaction coordinate does not vanish. Obviously, the summation of the energy components on all the mixed coordinates equals the energy component on the reaction coordinate along itself, which approximates the free energy along it. Thus, the energy decomposition scheme in Eq.~(\ref{eq:ener:comp}) is an approximate decomposition of the free energy along the reaction coordinate into components along each coordinate in a set of configurational coordinates of the system. 

\subsubsection*{2. When $\xi$ is a mixed coordinate}
In practice, the one-dimensional coordinate obtained can be just an approximation of the reaction coordinate, we thus assume that $\xi$ is a function of the reaction coordinate $q_{\rm s}$ and a bath mode $\eta_{\rm 1}$ without loss of generality. It is easy to show that the energy component on a bath mode along $\xi$ vanishes. Now, we consider the energy component on $z_{\rm i}$ that is a combination of the reaction coordinate $q_{\rm s}$ and a bath mode $\eta_{\rm 2}$, which can be the same as $\eta_{\rm 1}$ or be different to it, along $\xi$

\begin{subequations}
\begin{align}
\frac{dV_{z_{\rm i},\xi}}{d\xi}|_{\xi(\mvec{x})=\xi'} &= \frac{1}{F^{U}(\xi')}\int_{0}^{T}\frac{\partial V(\mvec{z}(t))}{\partial z_{\rm i}}[\frac{\partial z_{\rm i}}{\partial q_{\rm s}}\dot{q}_{\rm s}(t)+\frac{\partial z_{\rm i}}{\partial \eta_{\rm 2}}\dot{\eta}_{\rm 2}(t)]\delta(\xi(\mvec{x}(t))-\xi')h_{U}(t)dt \\
&\approx \langle \frac{\partial V(\mvec{z}(t))}{\partial z_{\rm i}}\frac{\partial z_{\rm i}}{\partial q_{\rm s}}\frac{\dot{q}_{\rm s}(t)}{\dot{\xi}(t)} \rangle_{\xi(\mvec{x})=\xi',U}^{Flux}
\end{align}
\label{eq:comp:mix2}
\end{subequations}

Summing over the energy components on all the mixed coordinates, we have

\begin{equation}
\langle \frac{dV(\mvec{z}(t))}{dt}\frac{1}{\dot{\xi}(t)} \rangle_{\xi(\mvec{x})=\xi',U}^{Flux} \approx \langle \frac{\partial V(\mvec{z}(t))}{\partial q_{\rm s}}\frac{\dot{q}_{\rm s}(t)}{\dot{\xi}(t)} \rangle_{\xi(\mvec{x})=\xi',U}^{Flux}
\label{eq:ener:sum}
\end{equation}
Thus, the summation of the energy components on all the coordinates along $\xi$ does not equal the energy component on $\xi$ along itself when $\xi$ is a mixed coordinate as there exists at least another mixed coordinate on which the energy component along $\xi$ does not vanish. Actually, the energy decomposition scheme given in Eq.~(\ref{eq:ener:comp}) aims to decompose the free energy along the true reaction coordinate rather than the one along $\xi$. When $\xi$ is very close to the reaction coordinate, the energy components should be quite good approximations to the ones along the reaction coordinate.

\subsubsection*{3. Energy along $\xi$ in the plateau region of the flux} 
When $\xi$ is a good one-dimensional reaction coordinate and there exits a plateau region in the flux along it. Let us define that $\xi_{\rm C}$ and $\xi_{\rm D}$ are the starting point and end point of the plateau region, respectively. From Eqs.~(\ref{eq:fluxave:df2}) and (\ref{eq:ener:sum}) with several transformations, the weighted average energy as a function of $\xi$ in the plateau region is

\begin{subequations}
\begin{align}
\langle V(\mvec{x}) \rangle_{\xi(\mvec{x})=\xi_{\rm D},U}^{Flux}-\langle V(\mvec{x}) \rangle_{\xi(\mvec{x})=\xi_{\rm C},U}^{Flux} &\approx \int_{\xi_{\rm C}}^{\xi_{\rm D}} \langle \frac{\partial V(\mvec{z}(t))}{\partial q_{\rm s}}\frac{\dot{q}_{\rm s}(t)}{\dot{\xi}(t)} \rangle_{\xi(\mvec{x})=\xi',U}^{Flux} d\xi' \\
&= \frac{1}{N_{\rm T}} \int_{0}^{T}\frac{\partial V(\mvec{z}(t))}{\partial q_{\rm s}}\dot{q}_{\rm s}(t)\int_{\xi_{\rm C}}^{\xi_{\rm D}}\delta(\xi(t)-\xi') d\xi'dt
\end{align}
\label{eq:comp:plateau}
\end{subequations}
From the above equation, it is now clear that the weighted average energy along $\xi$ is related to the energy component along the reaction coordinate rather than the one along $\xi$.

Now we formally define two free energy analogues $\breve{A}(\xi)$ and $\tilde{A}(\xi)$ as follows,

\begin{subequations}
\begin{align}
\breve{A}(\xi')&\equiv \langle V(\mvec{z}) \rangle_{\xi(\mvec{z})=\xi',U}^{Flux} \\
\frac{d\tilde{A}(\xi)}{d\xi}|_{\xi(\mvec{z})=\xi'}&\equiv \langle \frac{dV(\mvec{z}(t))}{dt}\frac{1}{\dot{\xi}(\mvec{z}(t))} \rangle_{\xi(\mvec{z})=\xi',U}^{Flux}
\end{align}
\label{eq:FEana:def}
\end{subequations}
The two free energy analogues are equivalent in the plateau region of the flux along $\xi$ and can be significantly different in other regions. Actually, Eq.~(\ref{eq:fluxave:df2}) provides a convenient way in practice to estimate the quantities in Eq.~(\ref{eq:ener:sum}). The calculation of $\frac{\partial V(\mvec{z}(t))}{\partial q_{\rm s}}$ is non-trivial and the estimation of $\frac{dV(\mvec{z}(t))}{dt}$ using a finite differences method requires the trajectory to be saved at almost every integration step in MD simulations, while $\breve{A}(\xi)$ can be computed from typical MD trajectories, in which the position and velocity informations are stored with a large time interval to save disk space in practice. 

\subsection*{D. Connection with emergent potential energy}
The energy component in Eq.\,(\ref{eq:ener:comp}) can be related to the emergent potential energy proposed in Ref.~[\cite{Li:2016aa}]. The emergent potential energy on a coordinate $z_{\rm i}$ along $\xi$ is defined as

\begin{equation}
\frac{\langle \Delta V_{z_{\rm i}}(\mvec{z};\xi',\Delta\xi)\rangle}{\Delta \xi} = \langle \frac{\partial V(\mvec{z})}{\partial z_{\rm i}}\frac{\dot{z}_{\rm i}}{\dot{\xi}}\text{sgn}(\dot{\xi}) \rangle_{\xi(\mvec{z})=\xi',U}=\langle \frac{\partial V(\mvec{z})}{\partial z_{\rm i}}\frac{\dot{z}_{\rm i}}{|\dot{\xi}|} \rangle_{\xi(\mvec{z})=\xi',U}
\label{eq:EPE}
\end{equation}
where sgn$(x)$ is the sign function and $\langle \cdots \rangle_{\xi(\mvec{z})=\xi',U}$ stands for an ensemble average over the ensemble $U$ at $\xi(\mvec{z})=\xi'$, which is defined as

\begin{equation}
\langle f(t)\rangle_{\xi(\mvec{x})=\xi',U}\equiv\frac{\int_{0}^{T}f(t)\delta(\xi(\mvec{x}(t))-\xi') h_{U}(t)dt}{\int_{0}^{T}\delta(\xi(\mvec{x}(t))-\xi') h_{U}(t)dt}
\label{eq:notation}
\end{equation}

On the other hand, the energy component on $z_{\rm i}$ along $\xi$ in this work can be rewritten with similar notations as

\begin{equation}
\langle \frac{\partial V(\mvec{z}(t))}{\partial z_{\rm i}}\frac{\dot{z}_{\rm i}(t)}{\dot{\xi}(t)} \rangle_{\xi(\mvec{x})=\xi',U}^{Flux} =\frac{\langle \frac{\partial V(\mvec{z})}{\partial z_{\rm i}}\dot{z}_{\rm i} \rangle_{\xi(\mvec{z})=\xi',U}}{\langle \dot{\xi} \rangle_{\xi(\mvec{z})=\xi',U}} 
\label{eq:ener:comp2}
\end{equation}

Recall the definition of a transmission coefficient analogue suggested in Ref.~[\cite{li2022time}],

\begin{equation}
\check{\kappa}^{U}(\xi')= \frac{\int_{0}^{T}\dot{\xi}(\mvec{x}(t))\delta(\xi(\mvec{x}(t))-\xi')h_{U}(t)dt}{\int_{0}^{T}|\dot{\xi}(\mvec{x}(t))|\delta(\xi(\mvec{x}(t))-\xi')h_{U}(t)dt}=\frac{\langle \dot{\xi} \rangle_{\xi(\mvec{z})=\xi',U}}{\langle |\dot{\xi}| \rangle_{\xi(\mvec{z})=\xi',U}}
\label{eq:coef}
\end{equation}

The multiplication of the energy component in Eq.~(\ref{eq:ener:comp2}) and the transition coefficient analogue in Eq.~(\ref{eq:coef}) gives

\begin{equation}
\langle \frac{\partial V(\mvec{z}(t))}{\partial z_{\rm i}}\frac{\dot{z}_{\rm i}(t)}{\dot{\xi}(t)} \rangle_{\xi(\mvec{x})=\xi',U}^{Flux}*\check{\kappa}^{U}(\xi')=\frac{\langle \frac{\partial V(\mvec{z})}{\partial z_{\rm i}}\dot{z}_{\rm i} \rangle_{\xi(\mvec{z})=\xi',U}}{\langle |\dot{\xi}| \rangle_{\xi(\mvec{z})=\xi',U}}
\label{eq:ener:coef}
\end{equation}

When $\frac{\partial V(\mvec{z})}{\partial z_{\rm i}}\frac{\dot{z}_{\rm i}}{|\dot{\xi}|}$ and $|\dot{\xi}|$ are independent to each other in the ensemble $U$, that is,

\begin{equation}
\langle \frac{\partial V(\mvec{z})}{\partial z_{\rm i}}\dot{z}_{\rm i} \rangle_{\xi(\mvec{z})=\xi',U}=\langle \frac{\partial V(\mvec{z})}{\partial z_{\rm i}}\frac{\dot{z}_{\rm i}}{|\dot{\xi}|} \rangle_{\xi(\mvec{z})=\xi',U}\langle |\dot{\xi}| \rangle_{\xi(\mvec{z})=\xi',U}
\label{eq:coor}
\end{equation}
then the multiplication of the energy component and the transition coefficient analogue gives the emergent potential energy.

\subsection*{E. New quantity for reaction coordinate identification}
Although the emergent potential energy provide a rigorous way to quantify from an energetic viewpoint the relative importance of a coordinate for a rare transition when the TPE is available, the estimation of it suffers from numerical instabilities in the region where the probability density in the TPE is very low. Here, we proposed the following quantity to appraise the relevance of a coordinate to the reaction coordinate

\begin{subequations}
\begin{align}
&\frac{d\hat{A}^{U}(\xi)}{d\xi}|_{\xi(\mvec{x})=\xi'}\equiv\frac{d\tilde{A}^{U}(\xi)}{d\xi}|_{\xi(\mvec{x})=\xi'}*\frac{F^{U}(\xi')}{N_{\rm T}}=\frac{1}{N_{\rm T}}\int_{0}^{T}\frac{dV(\mvec{x}(t))}{dt}\delta(\xi(\mvec{x}(t))-\xi') h_{U}(t)dt \\
\Rightarrow & \hat{A}^{U}(\xi_{\rm max})-\hat{A}^{U}(\xi_{\rm min})=\int^{\xi_{\rm max}}_{\xi_{\rm min}} \frac{d\tilde{A}^{U}(\xi)}{d\xi}|_{\xi(\mvec{x})=\xi'} d\xi'= \frac{1}{N_{\rm T}}\int_{0}^{T}\frac{dV(\mvec{x}(t))}{dt} h_{U}(t)dt
\end{align}
\label{eq:totPot:coor}
\end{subequations}
where, $\xi_{\rm min}$ and $\xi_{\rm max}$ are the lower and upper bounds of $\xi$ in the ensemble $U$, respectively. When the flux $F^{U}(\xi)$ along $\xi$ is vanishingly small, the term $\frac{F^{U}(\xi')}{N_{\rm T}}$ will render $\frac{d\check{A}^{U}(\xi)}{d\xi}|_{\xi(\mvec{x})=\xi'}$ to be close to zero and this quantity is unlikely to suffer from numerical instability. Since the difference of $\hat{A}^{U}(\xi)$ at the two boundaries of any coordinate are the same, it will be convenient via visual inspection to compare $\hat{A}^{U}(\xi)$ along different coordinates. When $\xi$ is the reaction coordinate, $\tilde{A}^{U}(\xi)$ is reproduced. When $\xi$ is a bath mode,   

\begin{subequations}
\begin{align}
&\frac{d\hat{A}^{U}(\xi)}{d\xi}|_{\xi(\mvec{x})=\xi'}\approx \frac{T^{U}\rho^U(\xi')}{N_{\rm T}}\langle \frac{dV(\mvec{x}(t))}{dq_{\rm s}}\dot{q}_{\rm s}\rangle_{\xi(\mvec{x})=\xi',U} \\
& \text{with}\quad T^{U}\equiv\int_{0}^{T}h_{U}(t)dt, \quad \rho^U(\xi')\equiv\frac{\int_{0}^{T}\delta(\xi(\mvec{x}(t))-\xi')h_{U}(t)dt}{\int_{0}^{T}h_{U}(t)dt}
\end{align}
\label{eq:totPot:bath}
\end{subequations}
where $T^{U}$ is the accumulated time in the ensemble $U$ and $\rho^U(\xi')$ is the probability density at $\xi(\mvec{x})=\xi'$ in the ensemble $U$. It should be reasonable to assume that $\langle \frac{dV(\mvec{x}(t))}{dq_{\rm s}}\dot{q}_{\rm s}\rangle_{\xi(\mvec{x})=\xi',U}$ is a constant and then one immediately has

\begin{subequations}
\begin{align}
&\hat{A}^{U}(\xi)\approx \overline{\Delta V^{U}}\int_{\xi_{\rm min}}^{\xi}\rho^U(\xi')d\xi' \\
& \text{with}\quad \overline{\Delta V^{U}}\equiv \frac{1}{N_{\rm T}}\int_{0}^{T}\frac{dV(\mvec{x}(t))}{dt} h_{U}(t)dt
\end{align}
\label{eq:totPot:bath}
\end{subequations}

\subsection*{F. Energy weighted current and multidimensional energy surface in the space of collective variables}
It is interesting to note that the average energy weighted with current intensity along the reaction coordinate is directly related to the free energy along it, as indicated in Eqs.~(\ref{eq:fluxave:df2}) and (\ref{eq:comp:RC}). While the conditional canonical average of potential energy gives the internal energy, the entropic component of the free energy seems mainly arises from the curvature of the transition tube or from the shape change of transition tube intersecting with the isosurfaces of the reaction coordinate. It could be insightful in the configurational space to look at an energy weighted current, which is defined as

\begin{equation}
\mvec{J}_{V}(\mvec{x})\equiv\int_{0}^{T}V(\mvec{x}(t))\dot{\mvec{x}}(t)\delta(\mvec{x}(t)-\mvec{x})dt=V(\mvec{x})\mvec{J}(\mvec{x})
\label{eq:enecur:x}
\end{equation}
One immediately has,

\begin{equation}
\nabla \mvec{J}_{V}(\mvec{x})=\nabla V(\mvec{x})\cdot \mvec{J}(\mvec{x})+V(\mvec{x})\nabla\cdot\mvec{J}(\mvec{x})
\label{eq:enecur:x2}
\end{equation}
In case of a steady state, i.e., $\nabla\cdot\mvec{J}(\mvec{x})=0$, then

\begin{equation}
\nabla \mvec{J}_{V}(\mvec{x})=\mvec{J}(\mvec{x})\cdot\nabla V(\mvec{x})=\int_{0}^{T}\frac{dV(\mvec{x}(t))}{dt}\delta(\mvec{x}(t)-\mvec{x})dt
\label{eq:enecur:x3}
\end{equation}
Thus, the divergence of the energy weighted current equals the directional derivative of the potential energy in the direction of the reactive current for a steady state process. $\mvec{J}(\mvec{x})\cdot\nabla V(\mvec{x})$ can also be viewed as the flux intensity along the gradient of the potential energy and is weighted by the magnitude of the gradient.   

The energy weighted current associated with an ensemble $U$ is then defined as,
\begin{equation}
\mvec{J}^{U}_{V}(\mvec{x})\equiv\int_{0}^{T}V(\mvec{x}(t))\dot{\mvec{x}}(t)\delta(\mvec{x}(t)-\mvec{x}) h_{U}(t)dt
\label{eq:enecur:xu}
\end{equation}
The projection of $\mvec{J}^{U}_{V}(\mvec{x})$ to the subspace of CVs gives the energy weighted current in the CVs subspace (Appendix B),

\begin{equation}
\mvec{J}^{U}_{V}(\mvec{y})=\int_{0}^{T}V(\mvec{x}(t))\dot{\mvec{y}}(t)\delta(\mvec{y}(t)-\mvec{y}) h_{U}(t)dt
\label{eq:ener:cur}
\end{equation}
Inspired by Eq.~(\ref{eq:enecur:x}), we thus seek for an energy function $\breve{A}^{U}_{V}(\mvec{y})$ of $\mvec{y}$ (thus a multidimensional free energy analogue in the subspace of CVs) such that $\mvec{J}^{U}_{V}(\mvec{y})=\breve{A}^{U}_{V}(\mvec{y})\mvec{J}^{U}(\mvec{y})$. One immediately finds that
\begin{equation}
\breve{A}^{U}_{V}(\mvec{y})=\frac{\mvec{J}^{U}_{V}(\mvec{y})\cdot \mvec{J}^{U}(\mvec{y})}{\mvec{J}^{U}(\mvec{y})\cdot \mvec{J}^{U}(\mvec{y})}
\label{eq:FE:ana}
\end{equation}
and that in the one-dimensional case $\breve{A}^{U}_{V}(\xi')=\langle V(\mvec{x})\rangle_{\xi(\mvec{x})=\xi',U}^{Flux}$.

In analogue to Eq.~(\ref{eq:enecur:x3}), the directional derivative of the potential energy associated to an ensemble $U$ is defined as,
\begin{equation}
j_{\nabla}^{U}(\mvec{x})\equiv\mvec{J}^{U}(\mvec{x})\cdot\nabla V(\mvec{x})=\int_{0}^{T}\frac{dV(\mvec{x}(t))}{dt}\delta(\mvec{x}(t)-\mvec{x})h_{U}(t)dt
\label{eq:curdiv}
\end{equation}
The projection of $j_{\nabla}^{U}(\mvec{x})$ to the subspace of CVs gives,
\begin{equation}
j_{\nabla}^{U}(\mvec{y})=\int j_{\nabla}^{U}(\mvec{x})\delta(\mvec{y}(\mvec{x})-\mvec{y})d\mvec{x}=\int_{0}^{T}\frac{dV(\mvec{x}(t))}{dt}\delta(\mvec{y}(t)-\mvec{y})h_{U}(t)dt
\label{eq:curdiv:cv}
\end{equation}
As suggested in Eq.~(\ref{eq:enecur:x3}), we seek for an free energy analogue $\tilde{A}^{U}_{V}(\mvec{y})$ such that $j_{\nabla}^{U}(\mvec{y})=\nabla\tilde{A}^{U}(\mvec{y})\cdot\mvec{J}^{U}(\mvec{y})$. Let us define $\tilde{\mvec{g}}^{U}_{V}(\mvec{y})$ as the gradient of the free energy analogue $\tilde{A}(\mvec{y})$, i.e., $\tilde{\mvec{g}}^{U}_{V}(\mvec{y})\equiv\nabla\tilde{A}^{U}(\mvec{y})$. It is easy to show that in the one-dimensional case $\tilde{\mvec{g}}^{U}_{V}(\xi')=\langle \frac{V(\mvec{x})}{dt}\frac{1}{\dot{\xi}(\mvec{x}(t))}\rangle_{\xi(\mvec{x})=\xi',U}^{Flux}=\frac{d\tilde{A}^{U}(\xi)}{d\xi}|_{\xi(\mvec{x})=\xi'}$.

In the region of the subspace of the CVs where the divergence of the reactive current vanishes, that is to say that there is no source or sink in this region for the ensemble $U$, we find that (Appendix C)
\begin{equation}
j_{\nabla}^{U}(\mvec{y})=\nabla\cdot\mvec{J}^{U}_{V}(\mvec{y}) \quad \text{at $\mvec{y}$ where $\mvec{J}^{U}_{V}(\mvec{y})$ is divergence-free}
\label{eq:FEana:equal}
\end{equation}

\subsubsection*{1. The choice of subsystem}
For complex systems, it is not practical to explore the energetics of the whole system at the per-coordinate level. Usually, a proper subsystem is selected to reduce the dimension and the noisiness due to the large fluctuations in the environment or bath modes. For example, it is shown to be feasible and useful when only the residues close to the retinal were included in the analysis of residue-residue mutual works in rhodopsin~\cite{li2020residue}. Let us define that the subsystem is described by $\mvec{y}$, and the rest of the system is described by $\mvec{y}'$ (i.e., $\{\mvec{y},\mvec{y}'\}$ is another complete and independent set of configurational coordinates). In terms of $\{\mvec{y},\mvec{y}'\}$, the potential energy $V(\mvec{y},\mvec{y}')=V_{\mvec{y}}(\mvec{y})+V_{\mvec{y}'}(\mvec{y}')+V_{\rm coupl}(\mvec{y},\mvec{y}')$. $V_{\mvec{y}}(\mvec{y})$ and $V_{\mvec{y}'}(\mvec{y}')$ are the potential energy for the subsystem and its environment, respectively. $V_{\rm coupl}(\mvec{y},\mvec{y}')$ is the interaction energy between the subsystem and its environment. We thus also assume that the reaction coordinate is a function of $\mvec{y}$ and $\mvec{y}'$ are bath modes, whose behaviours resemble the ones in an equilibrium ensemble.

The energy weighted current in the subspace of $\mvec{y}$ can then be expressed as

\begin{subequations}
\begin{align}
\mvec{J}^{U}_{V}(\mvec{y})&=\int_{0}^{T}[V_{\mvec{y}}(\mvec{y}(t))+V_{\mvec{y}'}(\mvec{y}'(t))+V_{\rm coupl}(\mvec{y}(t),\mvec{y}'(t))]\dot{\mvec{y}}(t)\delta(\mvec{y}(t)-\mvec{y}) h_{U}(t)dt \\
&= [V_{\mvec{y}}(\mvec{y})+\overline{V_{\mvec{y}'}}]\mvec{J}^{U}(\mvec{y})+\int_{0}^{T}V_{\rm coupl}(\mvec{y}(t),\mvec{y}'(t))\dot{\mvec{y}}(t)\delta(\mvec{y}(t)-\mvec{y}) h_{U}(t)dt
\end{align}
\label{eq:subsys:JV}
\end{subequations} 
where, $\overline{V_{\mvec{y}'}}$ is the average of $V_{\mvec{y}'}(\mvec{y}')$ in the equilibrium ensemble and thus a constant, which can be ignored. Thus, $\mvec{J}^{U}_{V}(\mvec{y})$ can be calculated by ignoring the term $V_{\mvec{y}'}(\mvec{y}')$, which is generally large and very noisy. 

In addition, we can write $j_{\nabla}^{U}(\mvec{y})$ as

\begin{subequations}
\begin{align}
j_{\nabla}^{U}(\mvec{y})&=\int_{0}^{T}[\sum_{i}\frac{\partial V(\mvec{y}(t),\mvec{y}'(t))}{\partial y_{\rm i}}\dot{y}_{\rm i}(t)+\sum_{j}\frac{\partial V(\mvec{y}(t),\mvec{y}'(t))}{\partial y'_{\rm j}}\dot{y}'_{\rm j}(t)]\delta(\mvec{y}(t)-\mvec{y})h_{U}(t)dt \\
 &= \int_{0}^{T}\sum_{i}\frac{\partial V(\mvec{y}(t),\mvec{y}'(t))}{\partial y_{\rm i}}\dot{y}_{\rm i}(t)\delta(\mvec{y}(t)-\mvec{y})h_{U}(t)dt \\
 &=\sum_{i}[\frac{\int_{0}^{T}\frac{\partial V(\mvec{y}(t),\mvec{y}'(t))}{\partial y_{\rm i}}\dot{y}_{\rm i}(t)\delta(\mvec{y}(t)-\mvec{y}) h_{U}(t)dt}{\int_{0}^{T}\dot{y}_{\rm i}(t)\delta(\mvec{y}(t)-\mvec{y}) h_{U}(t)dt}\int_{0}^{T}\dot{y}_{\rm i}(t)\delta(\mvec{y}(t)-\mvec{y}) h_{U}(t)dt]
\end{align}
\label{eq:subsys:JF}
\end{subequations} 
Thus, for the calculation of $j_{\nabla}^{U}(\mvec{y})$, only the generalized forces on $\mvec{y}$ are required. Note that $j_{\nabla}^{U}(\mvec{y})=\tilde{\mvec{g}}^{U}_{V}(\mvec{y})\cdot\mvec{J}^{U}(\mvec{y})$, one natural solution for $\tilde{\mvec{g}}^{U}_{V}(\mvec{y})$ is given below with the $i$-th element of $\tilde{\mvec{g}}^{U}_{V}(\mvec{y})$ being provided by

\begin{equation}
\tilde{\mvec{g}}^{U}_{V,\rm i}(\mvec{y})=\frac{\int_{0}^{T}\frac{\partial V(\mvec{y}(t),\mvec{y}'(t))}{\partial y_{\rm i}}\dot{y}_{\rm i}(t)\delta(\mvec{y}(t)-\mvec{y}) h_{U}(t)dt}{\int_{0}^{T}\dot{y}_{\rm i}(t)\delta(\mvec{y}(t)-\mvec{y}) h_{U}(t)dt}
\label{eq:grad:ana2}
\end{equation}
Substituting $V(\mvec{y},\mvec{y}')$ with $V_{\mvec{y}}(\mvec{y})+V_{\mvec{y}'}(\mvec{y}')+V_{\rm coupl}(\mvec{y},\mvec{y}')$, we find that

\begin{equation}
\tilde{\mvec{g}}^{U}_{V,\rm i}(\mvec{y})=\frac{\partial V_{\mvec{y}}(\mvec{y})}{\partial y_{\rm i}}+\frac{\int_{0}^{T}\frac{\partial V_{\rm coupl}(\mvec{y}(t),\mvec{y}'(t))}{\partial y_{\rm i}}\dot{y}_{\rm i}(t)\delta(\mvec{y}(t)-\mvec{y}) h_{U}(t)dt}{\int_{0}^{T}\dot{y}_{\rm i}(t)\delta(\mvec{y}(t)-\mvec{y}) h_{U}(t)dt} 
\label{eq:subsys:JF3}
\end{equation}

Now, we assume that $V_{\rm coupl}(\mvec{y},\mvec{y}')=g(\mvec{y})m(\mvec{y}')$, $\tilde{\mvec{g}}^{U}_{V,\rm i}(\mvec{y})$ can then be write as
\begin{equation}
\tilde{\mvec{g}}^{U}_{V,\rm i}(\mvec{y})=\frac{\partial [V_{\mvec{y}}(\mvec{y})+g(\mvec{y})\overline{m(\mvec{y}')}]}{\partial y_{\rm i}}
\label{eq:subsys:gv}
\end{equation}
where, $\overline{m(\mvec{y}')}$ is the average of $m(\mvec{y}')$ in the equilibrium ensemble and thus a constant. Similarly, we find that from Eq.~(\ref{eq:subsys:JV}b)

\begin{equation}
\mvec{J}^{U}_{V}(\mvec{y})= [V_{\mvec{y}}(\mvec{y})+g(\mvec{y})\overline{m(\mvec{y}')}+\overline{V_{\mvec{y}'}}]\mvec{J}^{U}(\mvec{y})
\label{eq:subsys:JV2}
\end{equation}
From Eqs.\,(\ref{eq:FE:ana}) and (\ref{eq:subsys:JV2}), we have 

\begin{equation}
\breve{A}^{U}_{V}(\mvec{y})=V_{\mvec{y}}(\mvec{y})+g(\mvec{y})\overline{m(\mvec{y}')}+\overline{V_{\mvec{y}'}}
\label{eq:FE:approx}
\end{equation}
Obviously, when $V_{\rm coupl}(\mvec{y},\mvec{y}')$ can be expressed as the summation of functions with the same form of $g(\mvec{y})m(\mvec{y}')$, Eq.~(\ref{eq:FE:approx}) holds. Thus, under the assumption that the distribution of $\mvec{y}'$ at any position of $\mvec{y}$ resemble the one in the equilibrium ensemble, $\tilde{A}^{U}(\mvec{y})$ is equivalent to $\breve{A}^{U}(\mvec{y})$ (i.e., differ by a constant).

If the coupling between the subsystem and its environment is weak, that is $V_{\rm coupl}(\mvec{y},\mvec{y}')\ll V_{\mvec{y}}(\mvec{y})$, $\tilde{\mvec{g}}^{U}_{V,\rm i}(\mvec{y})$ and $\breve{A}^{U}(\mvec{y})$ can be further approximated as

\begin{subequations}
\begin{align}
\tilde{\mvec{g}}^{U}_{V,\rm i}(\mvec{y})&\approx\frac{\partial V_{\mvec{y}}(\mvec{y})}{\partial y_{\rm i}} \\
\breve{A}^{U}(\mvec{y})&\approx V_{\mvec{y}}(\mvec{y})
\end{align}
\label{eq:subsys:JF4}
\end{subequations}
Thus, the above derivations rationalize that the free energy and the energy components can be obtained from a proper subsystem in close approximation. If the $\nabla \tilde{A}^{U}(\mvec{y})$ obtained with Eq.~(\ref{eq:subsys:JF4}a) is close to the one with Eq.~(\ref{eq:subsys:JF3}), the subsystem could be considered a proper one that preserves the free energy of the system. The energy components obtained from this subsystem are thus good approximations of the ones from the entire system. 

\subsubsection*{2. Overdamped dynamics and parabolic energy barrier}
Assuming an overdamped dynamics and a parabolic barrier of the potential of mean force $A(\mvec{y})$ with the barrier top at $\mvec{y}=0$, that is $\beta A(\mvec{y})=\mvec{y}^{t}\mvec{H}\mvec{y}/2$, where $\beta=1/(k_{\rm B}T)$ and $H$ is the Hessian matrix. From Roux~\cite{roux2021string} in case of a position-independent diffusion matrix $\mvec{D}$, the committor near the saddle point can be approximated by $q(\mvec{y})\approx q(\mvec{y}\cdot\mvec{e})$ with $\mvec{e}$ being the unit vector parallel to $\nabla q(\mvec{y})$, and $\mvec{e}$ is the eigenvector of the only negative eigenvalue $\lambda$ of $\mvec{H}\mvec{D}$, i.e., $\mvec{H}\mvec{D}\mvec{e}=-\lambda \mvec{e}$, with Eq.~(\ref{eq:flux:OD}) we can write $j_{\nabla}^{U}(\mvec{y})$ as

\begin{subequations}
\begin{align}
j_{\nabla}^{U}(\mvec{y})&\propto e^{-\beta A(\mvec{y})}\mvec{D}\nabla q(\mvec{y})\cdot\nabla A(\mvec{y}) \\
&\propto e^{-\beta A(\mvec{y})}\mvec{y}^{t}\mvec{H}\mvec{D}\mvec{e}q'(\mvec{y}\cdot\mvec{e}) \\
&\propto e^{-\beta A(\mvec{y})}q'(q_{\rm s})q_{\rm s}
\end{align}
\label{eq:ODsys:jx}
\end{subequations}
here, $q_{\rm s}=\mvec{y}\cdot\mvec{e}$ is the reaction coordinate. The gradient of $j_{\nabla}^{U}(\mvec{y})$ can be expressed as

\begin{subequations}
\begin{align}
\nabla j_{\nabla}^{U}(\mvec{y})&\propto e^{-\beta A(\mvec{y})}q'(q_{\rm s})\mvec{e}+q_{\rm s}[e^{-\beta A(\mvec{y})}q''(q_{\rm s})\mvec{e}+e^{-\beta A(\mvec{y})}q'(q_{\rm s})\nabla (-\beta A(\mvec{y}))]\\
&\propto e^{-\beta A(\mvec{y})}q'(q_{\rm s})[\mvec{e}-\frac{\lambda}{D}q_{\rm s}^{2}\mvec{e}-q_{\rm s}\nabla (\beta A(\mvec{y}))]
\end{align}
\label{eq:ODsys:jxg}
\end{subequations}
where, the conclusion that $q''(q_{\rm s})=-\lambda q_{\rm s}q'(q_{\rm s})/D$ from Ref.~[\cite{roux2021string}] is used with $D\equiv \mvec{e}^{t}\mvec{D}\mvec{e}$. In the vicinity of the saddle point, i.e., $q_{\rm}\approx 0$, the vector $\nabla j_{\nabla}^{U}(\mvec{y})$ is parallel to $\mvec{e}$. Thus, the analysis of $\nabla j_{\nabla}^{U}(\mvec{y})$ provides a way to find $\mvec{e}$, the direction of the gradient of the committor near the saddle point. 

\subsubsection*{3. A slow reaction coordinate coupled with fast bath modes}
Now we consider a system that consists of a slow coordinate $q_{\rm s}$, which is also the only reaction coordinate, and a pool of fast bath modes $\mvec{q}_{\rm b}$. It is assumed that the dynamics along bath modes in a non-equilibrium ensemble $U$ resembles the one in equilibrium and the coupling between the bath modes and the reaction coordinate $q_{\rm s}$ is weak. Since the components in the bath modes are vanishingly small, we can approximate $j_{\nabla}^{U}(\mvec{y})$ with

\begin{subequations}
\begin{align}
j_{\nabla}^{U}(q_{\rm s},\mvec{q}_{\rm b})&\approx \int_{0}^{T}\frac{\partial A(q_{\rm s}(t),\mvec{q}_{\rm b}(t))}{\partial q_{\rm s}}\dot{q}_{\rm s}(t)\delta(q_{\rm s}(t)-q_{\rm s})\delta(\mvec{q}_{\rm b}(t)-\mvec{q}_{\rm b})h_{U}(t)dt \\
&=\overline{K(q_{\rm s};\mvec{q}_{\rm b})}\rho^{U}(\mvec{q}_{\rm b})T^{U} \\
&\text{with}\quad \overline{K(q_{\rm s};\mvec{q}_{\rm b})}\equiv \frac{\int_{0}^{T}\frac{\partial A(q_{\rm s}(t),\mvec{q}_{\rm b}(t))}{\partial q_{\rm s}}\dot{q}_{\rm s}(t)\delta(q_{\rm s}(t)-q_{\rm s})\delta(\mvec{q}_{\rm b}(t)-\mvec{q}_{\rm b})h_{U}(t)dt}{\int_{0}^{T}\delta(\mvec{q}_{\rm b}(t)-\mvec{q}_{\rm b})h_{U}(t)dt} \\
&\text{and} \quad \rho^{U}(\mvec{q}_{\rm b})=\frac{\int_{0}^{T}\delta(\mvec{q}_{\rm b}(t)-\mvec{q}_{\rm b})h_{U}(t)dt}{\int_{0}^{T}h_{U}(t)dt}
\end{align}
\label{eq:SFsys:jx}
\end{subequations}
If $\overline{K(q_{\rm s};\mvec{q}_{\rm b})}$ is independent to $\mvec{q}_{\rm b}$, then

\begin{equation}
 \int  j_{\nabla}^{U}(q_{\rm s},\mvec{q}_{\rm b}) d\mvec{q}_{\rm b}\approx\int \overline{K(q_{\rm s};\mvec{q}_{\rm b})}\rho^{U}(\mvec{q}_{\rm b})T^{U} d\mvec{q}_{\rm b}=\overline{K(q_{\rm s};\mvec{q}_{\rm b})}T^{U}
\label{eq:SFsys:jx2}
\end{equation}
On the other hand, 
\begin{subequations}
\begin{align}
 \int  j_{\nabla}^{U}(q_{\rm s},\mvec{q}_{\rm b}) d\mvec{q}_{\rm b}&= \int_{0}^{T}\frac{d A(q_{\rm s}(t),\mvec{q}_{\rm b}(t))}{dt}\delta(q_{\rm s}(t)-q_{\rm s})h_{U}(t)dt \\ 
&= \langle \frac{d A(q_{\rm s}(t),\mvec{q}_{\rm b}(t))}{dt}\frac{1}{\dot{q}_{\rm s}(t)} \rangle_{q_{\rm s}(t)=q_{\rm s}}^{Flux}F^{U}(q_{\rm s})=\frac{d\tilde{A}(q_{\rm s})}{d q_{\rm s}}F^{U}(q_{\rm s})
\end{align}
\label{eq:SFsys:jx3}
\end{subequations}
Thus, $\overline{K(q_{\rm s};\mvec{q}_{\rm b})}T^{U}\approx\frac{d\tilde{A}(q_{\rm s})}{d q_{\rm s}}F^{U}(q_{\rm s})$ and 

\begin{equation}
j_{\nabla}^{U}(q_{\rm s},\mvec{q}_{\rm b})\approx \frac{d\tilde{A}(q_{\rm s})}{d q_{\rm s}}F^{U}(q_{\rm s})\rho^{U}(\mvec{q}_{\rm b})
\label{eq:SFsys:jx4}
\end{equation}
The gradient of $j_{\nabla}^{U}(q_{\rm s},\mvec{q}_{\rm b})$ is

\begin{subequations}
\begin{align}
\frac{\partial j_{\nabla}^{U}(q_{\rm s},\mvec{q}_{\rm b})}{\partial q_{\rm s}}&\approx [\frac{d^{2}\tilde{A}(q_{\rm s})}{d q^{2}_{\rm s}}F^{U}(q_{\rm s})+\frac{d\tilde{A}(q_{\rm s})}{d q_{\rm s}}\frac{dF^{U}(q_{\rm s})}{d q_{\rm s}}]\rho^{U}(\mvec{q}_{\rm b}) \\
\nabla_{\mvec{q}_{\rm b}}j_{\nabla}^{U}(q_{\rm s},\mvec{q}_{\rm b})&\approx\frac{d\tilde{A}(q_{\rm s})}{d q_{\rm s}}F^{U}(q_{\rm s})\nabla_{\mvec{q}_{\rm b}}\rho^{U}(\mvec{q}_{\rm b})
\end{align}
\label{eq:jx:grad}
\end{subequations}
$\nabla_{\mvec{q}_{\rm b}}$ denotes the derivatives with respect to $\mvec{q}_{\rm b}$. Eq.~(\ref{eq:jx:grad}) suggests that the gradient of $j_{\nabla}^{U}(q_{\rm s},\mvec{q}_{\rm b})$ points the direction of the reaction coordinate when (1) $\frac{d\tilde{A}(q_{\rm s})}{d q_{\rm s}}=0$ or (2) $\nabla_{\mvec{q}_{\rm b}}\rho^{U}(\mvec{q}_{\rm b})=\mvec{0}$, a zero vector. One particular case when $\frac{d\tilde{A}(q_{\rm s})}{d q_{\rm s}}=0$ is at the barrier top of the free energy analogue ($\tilde{A}(q_{\rm s})$) along the reaction coordinate. It is the surface $q_{\rm s}=q_{\rm s}^{*}$ with $\frac{d\tilde{A}(q_{\rm s})}{d q_{\rm s}}|_{q_{\rm s}=q_{\rm s}^{*}}=0$, which could be the stochastic separatrix. At the surface $q_{\rm s}=q_{\rm s}^{*}$, the directions of all $\nabla j_{\nabla}^{U}(q_{\rm s},\mvec{q}_{\rm b})$ are parallel to the direction of the reaction coordinate $q_{\rm s}$ and the magnitude of $\nabla j_{\nabla}^{U}(q_{\rm s},\mvec{q}_{\rm b}))$ is proportional to $\rho^{U}(\mvec{q}_{\rm b})$. The point at which the magnitude of $\nabla j_{\nabla}^{U}(q_{\rm s},\mvec{q}_{\rm b})$ reaches its maximum corresponds to the saddle point. A special case when $\nabla_{\mvec{q}_{\rm b}}\rho^{U}(\mvec{q}_{\rm b})=\mvec{0}$ is a curve where $\rho^{U}(\mvec{q}_{\rm b})$ reaches its maximum, which corresponds to the minimum free energy path and passes the saddle point. Thus, the direction of the reaction coordinate along the minimum free energy path can be obtained via visual inspection of the vector field $\nabla j_{\nabla}^{U}(\mvec{y})$ in the subspace of CVs.

\subsection*{F. Time-lagged quantities}
Analogue to the time-lagged flux and current proposed in Ref.~[\cite{li2022time}], the extension of the above quantities to time-lagged ones is rather straightforward, for instance, by replace the generalized velocity $\dot{\mvec{y}}(t)$ with $\frac{\mvec{y}(t+\tau)-\mvec{y}(t)}{\tau}$. Here, $\tau$ is the lag time. For example, the ensemble average weighted with current intensity given in Eq.~[\ref{eq:fluxave:f}] can be extended to the ensemble average weighted by time-lagged current intensity, which is defined as

\begin{equation}
\langle f(\mvec{x}) \rangle_{\xi(\mvec{x})=\xi',U}^{Flux,\tau}\equiv\frac{\int_{0}^{T}f(\mvec{x}(t))\frac{\xi(\mvec{x}(t+\tau))-\xi(\mvec{x}(t))}{\tau}\delta(\xi(\mvec{x}(t))-\xi') h_{U}(t)dt}{\int_{0}^{T}\frac{\xi(\mvec{x}(t+\tau))-\xi(\mvec{x}(t))}{\tau}\delta(\xi(\mvec{x}(t))-\xi') h_{U}(t)dt}
\label{eq:fluxave:tl}
\end{equation}

The time-lagged version of energy decomposed onto a coordinate $z_{\rm i}$ along the coordinate $\xi$ in Eq.~[\ref{eq:ener:comp}] is

\begin{equation}
V_{z_{\rm i},\xi}^{\tau}=\int \langle \frac{\partial V(\mvec{z}(t))}{\partial z_{\rm i}}\frac{z_{\rm i}(t+\tau)-z_{\rm i}(t)}{\xi(t+\tau)-\xi(t)} \rangle_{\xi(\mvec{x})=\xi',U}^{Flux,\tau}d\xi'
\label{eq:ener:comp:tl}
\end{equation}

The time-lagged counterpart of $j_{\nabla}^{U}(\mvec{y})$ in Eq.~[\ref{eq:curdiv:cv}] is,
\begin{equation}
j_{\nabla}^{U}(\mvec{y},\tau)\equiv\int_{0}^{T}\frac{V(\mvec{x}(t+\tau))-V(\mvec{x}(t))}{\tau}\delta(\mvec{y}(t)-\mvec{y})h_{U}(t)dt
\label{eq:curdiv:cv:tl}
\end{equation}

Time-lagged backward quantities and time-lagged mean quantities can be analogously defined. It will be interesting to explore the properties and usefulness of these time-lagged quantities in analysing non-equilibrium ensembles of MD trajectories.

\section*{\RNum{3}. CONCLUDING REMARKS}
In a non-equilibrium ensemble, an important characteristics is that the reactive current in it is non-zero. Following this thinking path, we here propose an ensemble average of an arbitrary time series weighted with the reactive current intensity, which can be related to the conditional canonical average under proper approximations. When the quantity to be averaged is the potential energy, such a weighted ensemble average is demonstrated directly related to the free energy along the reaction coordinate. By decomposing the infinitesimal change of the potential energy into components on the coordinates that form an independent and complete set of configurational coordinates, a free energy decomposition scheme is established in Eq.~(\ref{eq:ener:comp}) and is named as the energy decomposition along reaction coordinate. Although the energy decomposition is better to be done along the exact reaction coordinate, the energy decomposition along a good one-dimensional reaction coordinate gives quantitatively equivalent results, which are proportional to the ones along the true reaction coordinate as indicated in Eq.~(\ref{eq:comp:plateau}b). In addition, the energy component on a coordinate along the reaction coordinate is suggested to be able to appraise the relevance of the coordinate to the reaction coordinate and thus can be applied in the selection and optimization of a one-dimensional reaction coordinate.

Since the ensemble average in Eq.~(\ref{eq:fluxave:f}) gives a free energy analogue as a function of $\xi$, one may ask whether a multi-dimensional free energy analogue can be defined based on the reactive current in the subspace of CVs? We thus proposed the potential energy weighted current and the directional derivative of the potential energy along the reactive current, from which free energy analogues $\tilde{A}(\mvec{y})$ and $\breve{A}(\mvec{y})$ are defined, respectively. These free energy analogues can be considered as the extension of the one-dimensional free energy analogue. Although $\tilde{A}(\mvec{y})$ and $\breve{A}(\mvec{y})$ are equivalent at regions where the reactive current is divergence-free, $\tilde{A}(\mvec{y})$ should be in a closer relation than $\breve{A}(\mvec{y})$ with the free energy at regions where the divergence of the reactive current does not vanish. From a practical aspect, we would like to emphasize that Eq.~(\ref{eq:FE:ana}) provides a convenient way to obtain a reasonable estimation of the multidimensional free energy in the region where the reactive current is divergence-free, while Eq.~(\ref{eq:grad:ana2}) can be generally applied to calculate the free energy surface from a non-equilibrium ensemble in which the reactive current is non-vanishing.

Interestingly, $j_{\nabla}^{U}(\mvec{y})$, the directional derivative of the free energy analogue $\tilde{A}(\mvec{y})$ as defined in Eq.~(\ref{eq:curdiv:cv}), could be more useful than it appears. The gradient of $j_{\nabla}^{U}(\mvec{y})$ gives a vector field from which the direction of the reaction coordinate can be easily identified along the minimum free energy path in the subspace of CVs (not the tangent of the minimum free energy path). We suggest that a few relevant coordinates can be first identified with Eq.~(\ref{eq:totPot:coor}a) without the calculation of partial derivative with respect to coordinates (an independent and complete set) and then a one-dimensional reaction coordinate as a (possibly curvilinear) function of these relevant coordinates can be constructed by visual inspection of the vector field obtained from $j_{\nabla}^{U}(\mvec{y})$ with Eq.~(\ref{eq:subsys:JF}b).

We would like to point out that the potential energy in Eqs.~(\ref{eq:ener:cur}) and (\ref{eq:curdiv:cv}) can be replaced by an arbitrary function of $\mvec{x}$ or more generally an arbitrary time series. Quantities analogous to $\tilde{A}(\mvec{y})$ and $\breve{A}(\mvec{y})$ can then be defined as well and a relation similar to the one in Eq.~(\ref{eq:FEana:equal}) can also be obtained.

About the choice of the non-equilibrium ensemble, a natural choice is the TPE, in which the reactive flux along the reaction coordinate is known to be equal to the number of the transition paths. Another interesting ensemble could be the swarms of short trajectories, for example, the short trajectories initiated from configurations in a chain of replicas as in the string method~\cite{pan2008finding} or a known transition path. One can imaging that the flux along the reaction coordinate in the transition state region will be close to zero as the forward (go towards the product state) and backward (go towards the reactant state) trajectories are equally probable. In the barrier region close to the reactant state, there will be more backward trajectory than the forward trajectory and thus the net flux will be negative, while the flux will be positive in the barrier region close to the product state. Thus, the quantities proposed here can generally be estimated for such an ensemble of trajectories. Alternatively, one can reverse the backward trajectories and thus convert them into forward trajectories and reweight these trajectories (although it is not trivial), then there will be a net positive flux along the reaction coordinate (including the transition state region) and the above quantities can be generally evaluated to gain insightful information about the transition process. 

\section*{APPENDIX}
\subsection*{(A) Proof of Eq.~(\ref{eq:fluxave:df2})}
In the plateau region of the flux ($F^{U}(\xi)$) along $\xi$, we have that $F^{U}(\xi)$ equals the number of transition paths $N_{\rm T}$ and $\frac{dF^{U}(\xi)}{d\xi}|_{\xi=\xi'}=0$.  Note that $\langle f(\mvec{x}) \rangle_{\xi(\mvec{x})=\xi',U}^{Flux}$ in Eq.\,(\ref{eq:fluxave:f}) is a function of $\xi'$, we can consider its derivative with respect to $\xi'$ for $\xi'$ in the plateau region,

\begin{subequations}
\begin{align}
\frac{d\langle f(\mvec{x}) \rangle_{\xi(\mvec{x})=\xi',U}^{Flux}}{d\xi'}=&\frac{1}{N_{\rm T}}\int_{0}^{T}f(\mvec{x}(t))\dot{\xi}(\mvec{x}(t))\frac{d\delta(\xi(\mvec{x}(t))-\xi')}{d\xi'} h_{U}(t)dt  \\ 
=& \frac{1}{N_{\rm T}}\int_{0}^{T}f(\mvec{x}(t))\frac{-d\delta(\xi(\mvec{x}(t))-\xi')}{dt} h_{U}(t)dt \\
=& \frac{1}{N_{\rm T}}\int_{0}^{T}\frac{d[f(\mvec{x}(t))h_{U}(t)]}{dt}\delta(\xi(\mvec{x}(t))-\xi')dt \\
=& \langle \frac{df(\mvec{x}(t))}{dt}\frac{1}{\dot{\xi}(\mvec{x}(t))} \rangle_{\xi(\mvec{x})=\xi',U}^{Flux} 
\end{align}
\label{eq:fluxave:df}
\end{subequations}
From Eq.~(\ref{eq:fluxave:df}b) to Eq.~(\ref{eq:fluxave:df}c), integration by parts is used and the fact is noted that there is no source or sink in the plateau region as indicated in $\frac{dF^{U}(\xi)}{d\xi}|_{\xi=\xi'}=0$. From Eq.~(\ref{eq:fluxave:df}c) to Eq.~(\ref{eq:fluxave:df}d), we note that $dh_{U}(t)/dt=0$ for any time when the trajectory hits the surface $\xi(\mvec{x})=\xi'$.

\subsection*{(B) The energy weighted current in the subspace of collective variables}
The $i$-th element of $\mvec{J}^{U}_{V}(\mvec{y})$ is the projection of $\mvec{J}^{U}_{V}(\mvec{x})$ to $y_{\rm i}$ in the subspace of CVs, that is

\begin{subequations}
\begin{align}
\mvec{J}^{U}_{V,\rm i}(\mvec{y})&=\int \mvec{J}^{U}_{V}(\mvec{x})\nabla y_{\rm i}(\mvec{x}) \delta(\mvec{y}(\mvec{x})-\mvec{y})d\mvec{x} \\
&=\int_{0}^{T}V(\mvec{x}(t))\dot{\mvec{x}}(t)\cdot\int\nabla y_{\rm i}(\mvec{x})\delta(\mvec{x}(t)-\mvec{x})\delta(\mvec{y}(\mvec{x})-\mvec{y})d\mvec{x}  h_{U}(t)dt \\
&=\int_{0}^{T}V(\mvec{x}(t))\nabla y_{\rm i}(\mvec{x}(t))\cdot \dot{\mvec{x}}(t)\delta(\mvec{y}(\mvec{x}(t))-\mvec{y}) h_{U}(t)dt \\
&=\int_{0}^{T}V(\mvec{x}(t))\dot{y}_{\rm i}(\mvec{x}(t))\delta(\mvec{y}(\mvec{x}(t))-\mvec{y}) h_{U}(t)dt
\end{align}
\label{eq:ener:cur:X}
\end{subequations}

\subsection*{(C) Proof of Eq.~(\ref{eq:FEana:equal})}
In the region of the subspace of the CVs where the divergence of the reactive current vanishes, i.e., $\nabla\cdot\mvec{J}^{U}(\mvec{y})=0$, which indicates that there is no source or sink in this region for the ensemble $U$, we find that
\begin{subequations}
\begin{align}
j_{\nabla}^{U}(\mvec{y})&=\int_{0}^{T}\frac{dV(\mvec{x}(t))}{dt}\delta(\mvec{y}(t)-\mvec{y})h_{U}(t)dt \\
&= - \int_{0}^{T}\frac{d[\delta(\mvec{y}(t)-\mvec{y})h_{U}(t)]}{dt}V(\mvec{x}(t))dt \\
&= - \int_{0}^{T}\frac{d\delta(\mvec{y}(t)-\mvec{y})}{dt}V(\mvec{x}(t))h_{U}(t)dt \\
&=\int_{0}^{T}V(\mvec{x}(t))\dot{\mvec{y}}(t)\cdot\nabla\delta(\mvec{y}(t)-\mvec{y}) h_{U}(t)dt \\
&=\nabla\cdot\mvec{J}^{U}_{V}(\mvec{y})
\end{align}
\label{eq:appendix:c}
\end{subequations}
From Eq.~(\ref{eq:appendix:c}a) to Eq.~(\ref{eq:appendix:c}b), integration by parts is applied. From Eq.~(\ref{eq:appendix:c}b) to Eq.~(\ref{eq:appendix:c}c), we note that $dh_{U}(t)/dt=0$ for any time when the trajectory hits the surface $\mvec{y}(\mvec{x})=\mvec{y}$.

\section*{ACKNOWLEDGMENTS}
This work was supported by Natural Science Foundation of Guangdong Province, China (Grant No. 2020A1515010984) and the Start-up Grant for Young Scientists (860-000002110384), Shenzhen University.

\bibliography{./references-flux}

\providecommand{\latin}[1]{#1}
\makeatletter
\providecommand{\doi}
  {\begingroup\let\do\@makeother\dospecials
  \catcode`\{=1 \catcode`\}=2 \doi@aux}
\providecommand{\doi@aux}[1]{\endgroup\texttt{#1}}
\makeatother
\providecommand*\mcitethebibliography{\thebibliography}
\csname @ifundefined\endcsname{endmcitethebibliography}
  {\let\endmcitethebibliography\endthebibliography}{}
\begin{mcitethebibliography}{39}
\providecommand*\natexlab[1]{#1}
\providecommand*\mciteSetBstSublistMode[1]{}
\providecommand*\mciteSetBstMaxWidthForm[2]{}
\providecommand*\mciteBstWouldAddEndPuncttrue
  {\def\EndOfBibitem{\unskip.}}
\providecommand*\mciteBstWouldAddEndPunctfalse
  {\let\EndOfBibitem\relax}
\providecommand*\mciteSetBstMidEndSepPunct[3]{}
\providecommand*\mciteSetBstSublistLabelBeginEnd[3]{}
\providecommand*\EndOfBibitem{}
\mciteSetBstSublistMode{f}
\mciteSetBstMaxWidthForm{subitem}{(\alph{mcitesubitemcount})}
\mciteSetBstSublistLabelBeginEnd
  {\mcitemaxwidthsubitemform\space}
  {\relax}
  {\relax}

\bibitem[Karplus and McCammon(2002)Karplus, and McCammon]{karplus2002molecular}
Karplus,~M.; McCammon,~J.~A. \emph{Nature structural biology} \textbf{2002},
  \emph{9}, 646--652\relax
\mciteBstWouldAddEndPuncttrue
\mciteSetBstMidEndSepPunct{\mcitedefaultmidpunct}
{\mcitedefaultendpunct}{\mcitedefaultseppunct}\relax
\EndOfBibitem
\bibitem[Wang \latin{et~al.}(2021)Wang, Singh, and Li]{wang2021molecular}
Wang,~X.; Singh,~N.; Li,~W. \emph{Systems Medicine}; Elsevier, 2021; pp
  182--189\relax
\mciteBstWouldAddEndPuncttrue
\mciteSetBstMidEndSepPunct{\mcitedefaultmidpunct}
{\mcitedefaultendpunct}{\mcitedefaultseppunct}\relax
\EndOfBibitem
\bibitem[Chandler(1978)]{Chandler:1978aa}
Chandler,~D. \emph{The Journal of Chemical Physics} \textbf{1978}, \emph{68},
  2959--2970\relax
\mciteBstWouldAddEndPuncttrue
\mciteSetBstMidEndSepPunct{\mcitedefaultmidpunct}
{\mcitedefaultendpunct}{\mcitedefaultseppunct}\relax
\EndOfBibitem
\bibitem[Pan \latin{et~al.}(2008)Pan, Sezer, and Roux]{pan2008finding}
Pan,~A.~C.; Sezer,~D.; Roux,~B. \emph{The journal of physical chemistry B}
  \textbf{2008}, \emph{112}, 3432--3440\relax
\mciteBstWouldAddEndPuncttrue
\mciteSetBstMidEndSepPunct{\mcitedefaultmidpunct}
{\mcitedefaultendpunct}{\mcitedefaultseppunct}\relax
\EndOfBibitem
\bibitem[Allen \latin{et~al.}(2009)Allen, Valeriani, and ten
  Wolde]{Allen:2009aa}
Allen,~R.~J.; Valeriani,~C.; ten Wolde,~P.~R. \emph{Journal of Physics:
  Condensed Matter} \textbf{2009}, \emph{21}, 463102\relax
\mciteBstWouldAddEndPuncttrue
\mciteSetBstMidEndSepPunct{\mcitedefaultmidpunct}
{\mcitedefaultendpunct}{\mcitedefaultseppunct}\relax
\EndOfBibitem
\bibitem[Borrero and Escobedo(2007)Borrero, and Escobedo]{Borrero:2007aa}
Borrero,~E.~E.; Escobedo,~F.~A. \emph{The Journal of chemical physics}
  \textbf{2007}, \emph{127}, 164101\relax
\mciteBstWouldAddEndPuncttrue
\mciteSetBstMidEndSepPunct{\mcitedefaultmidpunct}
{\mcitedefaultendpunct}{\mcitedefaultseppunct}\relax
\EndOfBibitem
\bibitem[Faradjian and Elber(2004)Faradjian, and Elber]{faradjian2004computing}
Faradjian,~A.~K.; Elber,~R. \emph{The Journal of chemical physics}
  \textbf{2004}, \emph{120}, 10880--10889\relax
\mciteBstWouldAddEndPuncttrue
\mciteSetBstMidEndSepPunct{\mcitedefaultmidpunct}
{\mcitedefaultendpunct}{\mcitedefaultseppunct}\relax
\EndOfBibitem
\bibitem[Elber(2017)]{elber2017new}
Elber,~R. \emph{Quarterly reviews of biophysics} \textbf{2017}, \emph{50}\relax
\mciteBstWouldAddEndPuncttrue
\mciteSetBstMidEndSepPunct{\mcitedefaultmidpunct}
{\mcitedefaultendpunct}{\mcitedefaultseppunct}\relax
\EndOfBibitem
\bibitem[Berezhkovskii and Szabo(2019)Berezhkovskii, and
  Szabo]{berezhkovskii2019committors}
Berezhkovskii,~A.~M.; Szabo,~A. \emph{The Journal of chemical physics}
  \textbf{2019}, \emph{150}, 054106\relax
\mciteBstWouldAddEndPuncttrue
\mciteSetBstMidEndSepPunct{\mcitedefaultmidpunct}
{\mcitedefaultendpunct}{\mcitedefaultseppunct}\relax
\EndOfBibitem
\bibitem[Zuckerman and Chong(2017)Zuckerman, and Chong]{zuckerman2017weighted}
Zuckerman,~D.~M.; Chong,~L.~T. \emph{Annual review of biophysics}
  \textbf{2017}, \emph{46}, 43--57\relax
\mciteBstWouldAddEndPuncttrue
\mciteSetBstMidEndSepPunct{\mcitedefaultmidpunct}
{\mcitedefaultendpunct}{\mcitedefaultseppunct}\relax
\EndOfBibitem
\bibitem[Dellago \latin{et~al.}(1998)Dellago, Bolhuis, Csajka, and
  Chandler]{Dellago:1998aa}
Dellago,~C.; Bolhuis,~P.~G.; Csajka,~F.~S.; Chandler,~D. \emph{The Journal of
  Chemical Physics} \textbf{1998}, \emph{108}, 1964\relax
\mciteBstWouldAddEndPuncttrue
\mciteSetBstMidEndSepPunct{\mcitedefaultmidpunct}
{\mcitedefaultendpunct}{\mcitedefaultseppunct}\relax
\EndOfBibitem
\bibitem[Bolhuis \latin{et~al.}(1998)Bolhuis, Dellago, and
  Chandler]{Bolhuis:1998aa}
Bolhuis,~P.~G.; Dellago,~C.; Chandler,~D. \emph{Faraday Discussions}
  \textbf{1998}, \emph{110}, 421--436\relax
\mciteBstWouldAddEndPuncttrue
\mciteSetBstMidEndSepPunct{\mcitedefaultmidpunct}
{\mcitedefaultendpunct}{\mcitedefaultseppunct}\relax
\EndOfBibitem
\bibitem[Dellago \latin{et~al.}(1999)Dellago, Bolhuis, and
  Chandler]{Dellago:1999aa}
Dellago,~C.; Bolhuis,~P.~G.; Chandler,~D. \emph{The Journal of Chemical
  Physics} \textbf{1999}, \emph{110}, 6617\relax
\mciteBstWouldAddEndPuncttrue
\mciteSetBstMidEndSepPunct{\mcitedefaultmidpunct}
{\mcitedefaultendpunct}{\mcitedefaultseppunct}\relax
\EndOfBibitem
\bibitem[Bolhuis \latin{et~al.}(2002)Bolhuis, Chandler, Dellago, and
  Geissler]{Bolhuis:2002aa}
Bolhuis,~P.~G.; Chandler,~D.; Dellago,~C.; Geissler,~P.~L. \emph{Annual Review
  of Physical Chemistry} \textbf{2002}, \emph{53}, 291--318\relax
\mciteBstWouldAddEndPuncttrue
\mciteSetBstMidEndSepPunct{\mcitedefaultmidpunct}
{\mcitedefaultendpunct}{\mcitedefaultseppunct}\relax
\EndOfBibitem
\bibitem[Dellago \latin{et~al.}(2002)Dellago, Bolhuis, and
  Geissler]{Dellago:2002aa}
Dellago,~C.; Bolhuis,~P.~G.; Geissler,~P.~L. \emph{Adv. in Chem. Phys.}
  \textbf{2002}, \emph{123}, 1--78\relax
\mciteBstWouldAddEndPuncttrue
\mciteSetBstMidEndSepPunct{\mcitedefaultmidpunct}
{\mcitedefaultendpunct}{\mcitedefaultseppunct}\relax
\EndOfBibitem
\bibitem[Roux(2021)]{roux2021string}
Roux,~B. \emph{The Journal of Physical Chemistry A} \textbf{2021}, \emph{125},
  7558--7571\relax
\mciteBstWouldAddEndPuncttrue
\mciteSetBstMidEndSepPunct{\mcitedefaultmidpunct}
{\mcitedefaultendpunct}{\mcitedefaultseppunct}\relax
\EndOfBibitem
\bibitem[Li and Ma(2014)Li, and Ma]{li2014recent}
Li,~W.; Ma,~A. \emph{Molecular simulation} \textbf{2014}, \emph{40},
  784--793\relax
\mciteBstWouldAddEndPuncttrue
\mciteSetBstMidEndSepPunct{\mcitedefaultmidpunct}
{\mcitedefaultendpunct}{\mcitedefaultseppunct}\relax
\EndOfBibitem
\bibitem[Berezhkovskii and Szabo(2005)Berezhkovskii, and
  Szabo]{berezhkovskii2005one}
Berezhkovskii,~A.; Szabo,~A. \emph{The Journal of chemical physics}
  \textbf{2005}, \emph{122}, 014503\relax
\mciteBstWouldAddEndPuncttrue
\mciteSetBstMidEndSepPunct{\mcitedefaultmidpunct}
{\mcitedefaultendpunct}{\mcitedefaultseppunct}\relax
\EndOfBibitem
\bibitem[Rhee and Pande(2005)Rhee, and Pande]{rhee2005one}
Rhee,~Y.~M.; Pande,~V.~S. \emph{The Journal of Physical Chemistry B}
  \textbf{2005}, \emph{109}, 6780--6786\relax
\mciteBstWouldAddEndPuncttrue
\mciteSetBstMidEndSepPunct{\mcitedefaultmidpunct}
{\mcitedefaultendpunct}{\mcitedefaultseppunct}\relax
\EndOfBibitem
\bibitem[Ma and Dinner(2005)Ma, and Dinner]{Ma:2005aa}
Ma,~A.; Dinner,~A.~R. \emph{J Phys Chem B} \textbf{2005}, \emph{109},
  6769--79\relax
\mciteBstWouldAddEndPuncttrue
\mciteSetBstMidEndSepPunct{\mcitedefaultmidpunct}
{\mcitedefaultendpunct}{\mcitedefaultseppunct}\relax
\EndOfBibitem
\bibitem[Peters and Trout(2006)Peters, and Trout]{Peters:2006aa}
Peters,~B.; Trout,~B.~L. \emph{The Journal of Chemical Physics} \textbf{2006},
  \emph{125}, 054108\relax
\mciteBstWouldAddEndPuncttrue
\mciteSetBstMidEndSepPunct{\mcitedefaultmidpunct}
{\mcitedefaultendpunct}{\mcitedefaultseppunct}\relax
\EndOfBibitem
\bibitem[Peters \latin{et~al.}(2007)Peters, Beckham, and Trout]{Peters:2007aa}
Peters,~B.; Beckham,~G.~T.; Trout,~B.~L. \emph{The Journal of chemical physics}
  \textbf{2007}, \emph{127}, 034109--034109\relax
\mciteBstWouldAddEndPuncttrue
\mciteSetBstMidEndSepPunct{\mcitedefaultmidpunct}
{\mcitedefaultendpunct}{\mcitedefaultseppunct}\relax
\EndOfBibitem
\bibitem[Peters \latin{et~al.}(2013)Peters, Bolhuis, Mullen, and
  Shea]{Peters:2013aa}
Peters,~B.; Bolhuis,~P.~G.; Mullen,~R.~G.; Shea,~J.-E. \emph{The Journal of
  chemical physics} \textbf{2013}, \emph{138}, 054106\relax
\mciteBstWouldAddEndPuncttrue
\mciteSetBstMidEndSepPunct{\mcitedefaultmidpunct}
{\mcitedefaultendpunct}{\mcitedefaultseppunct}\relax
\EndOfBibitem
\bibitem[Li and Ma(2015)Li, and Ma]{li2015reducing}
Li,~W.; Ma,~A. \emph{The Journal of chemical physics} \textbf{2015},
  \emph{143}, 11B603\_1\relax
\mciteBstWouldAddEndPuncttrue
\mciteSetBstMidEndSepPunct{\mcitedefaultmidpunct}
{\mcitedefaultendpunct}{\mcitedefaultseppunct}\relax
\EndOfBibitem
\bibitem[Li and Ma(2016)Li, and Ma]{Li:2016aa}
Li,~W.; Ma,~A. \emph{The Journal of chemical physics} \textbf{2016},
  \emph{144}, 134104\relax
\mciteBstWouldAddEndPuncttrue
\mciteSetBstMidEndSepPunct{\mcitedefaultmidpunct}
{\mcitedefaultendpunct}{\mcitedefaultseppunct}\relax
\EndOfBibitem
\bibitem[Li(2018)]{li2018equipartition}
Li,~W. \emph{The Journal of chemical physics} \textbf{2018}, \emph{148},
  084105\relax
\mciteBstWouldAddEndPuncttrue
\mciteSetBstMidEndSepPunct{\mcitedefaultmidpunct}
{\mcitedefaultendpunct}{\mcitedefaultseppunct}\relax
\EndOfBibitem
\bibitem[Li(2022)]{li2021optimizing}
Li,~W. \emph{The Journal of Chemical Physics} \textbf{2022}, \emph{156},
  054117\relax
\mciteBstWouldAddEndPuncttrue
\mciteSetBstMidEndSepPunct{\mcitedefaultmidpunct}
{\mcitedefaultendpunct}{\mcitedefaultseppunct}\relax
\EndOfBibitem
\bibitem[Li(2022)]{li2022time}
Li,~W. \emph{bioRxiv} \textbf{2022},
  \emph{doi.org/10.1101/2022.02.23.481712}\relax
\mciteBstWouldAddEndPuncttrue
\mciteSetBstMidEndSepPunct{\mcitedefaultmidpunct}
{\mcitedefaultendpunct}{\mcitedefaultseppunct}\relax
\EndOfBibitem
\bibitem[Boresch \latin{et~al.}(1994)Boresch, Archontis, and
  Karplus]{boresch1994free}
Boresch,~S.; Archontis,~G.; Karplus,~M. \emph{Proteins: Structure, Function,
  and Bioinformatics} \textbf{1994}, \emph{20}, 25--33\relax
\mciteBstWouldAddEndPuncttrue
\mciteSetBstMidEndSepPunct{\mcitedefaultmidpunct}
{\mcitedefaultendpunct}{\mcitedefaultseppunct}\relax
\EndOfBibitem
\bibitem[Brady and Sharp(1995)Brady, and Sharp]{brady1995decomposition}
Brady,~G.~P.; Sharp,~K.~A. \emph{Journal of molecular biology} \textbf{1995},
  \emph{254}, 77--85\relax
\mciteBstWouldAddEndPuncttrue
\mciteSetBstMidEndSepPunct{\mcitedefaultmidpunct}
{\mcitedefaultendpunct}{\mcitedefaultseppunct}\relax
\EndOfBibitem
\bibitem[Srinivasan \latin{et~al.}(1998)Srinivasan, Cheatham, Cieplak, Kollman,
  and Case]{srinivasan1998continuum}
Srinivasan,~J.; Cheatham,~T.~E.; Cieplak,~P.; Kollman,~P.~A.; Case,~D.~A.
  \emph{Journal of the American Chemical Society} \textbf{1998}, \emph{120},
  9401--9409\relax
\mciteBstWouldAddEndPuncttrue
\mciteSetBstMidEndSepPunct{\mcitedefaultmidpunct}
{\mcitedefaultendpunct}{\mcitedefaultseppunct}\relax
\EndOfBibitem
\bibitem[Gohlke and Case(2004)Gohlke, and Case]{gohlke2004converging}
Gohlke,~H.; Case,~D.~A. \emph{Journal of computational chemistry}
  \textbf{2004}, \emph{25}, 238--250\relax
\mciteBstWouldAddEndPuncttrue
\mciteSetBstMidEndSepPunct{\mcitedefaultmidpunct}
{\mcitedefaultendpunct}{\mcitedefaultseppunct}\relax
\EndOfBibitem
\bibitem[Smith and van Gunsteren(1994)Smith, and van Gunsteren]{smith1994free}
Smith,~P.~E.; van Gunsteren,~W.~F. \emph{The Journal of Physical Chemistry}
  \textbf{1994}, \emph{98}, 13735--13740\relax
\mciteBstWouldAddEndPuncttrue
\mciteSetBstMidEndSepPunct{\mcitedefaultmidpunct}
{\mcitedefaultendpunct}{\mcitedefaultseppunct}\relax
\EndOfBibitem
\bibitem[Li(2020)]{li2020residue}
Li,~W. \emph{Journal of Chemical Theory and Computation} \textbf{2020},
  \emph{16}, 1834--1842\relax
\mciteBstWouldAddEndPuncttrue
\mciteSetBstMidEndSepPunct{\mcitedefaultmidpunct}
{\mcitedefaultendpunct}{\mcitedefaultseppunct}\relax
\EndOfBibitem
\bibitem[Weinan and Vanden-Eijnden(2010)Weinan, and
  Vanden-Eijnden]{vanden2010transition}
Weinan,~E.; Vanden-Eijnden,~E. \emph{Annual review of physical chemistry}
  \textbf{2010}, \emph{61}, 391--420\relax
\mciteBstWouldAddEndPuncttrue
\mciteSetBstMidEndSepPunct{\mcitedefaultmidpunct}
{\mcitedefaultendpunct}{\mcitedefaultseppunct}\relax
\EndOfBibitem
\bibitem[Johnson and Hummer(2012)Johnson, and
  Hummer]{johnson2012characterization}
Johnson,~M.~E.; Hummer,~G. \emph{The Journal of Physical Chemistry B}
  \textbf{2012}, \emph{116}, 8573--8583\relax
\mciteBstWouldAddEndPuncttrue
\mciteSetBstMidEndSepPunct{\mcitedefaultmidpunct}
{\mcitedefaultendpunct}{\mcitedefaultseppunct}\relax
\EndOfBibitem
\bibitem[Zwanzig(1973)]{zwanzig1973nonlinear}
Zwanzig,~R. \emph{Journal of Statistical Physics} \textbf{1973}, \emph{9},
  215--220\relax
\mciteBstWouldAddEndPuncttrue
\mciteSetBstMidEndSepPunct{\mcitedefaultmidpunct}
{\mcitedefaultendpunct}{\mcitedefaultseppunct}\relax
\EndOfBibitem
\bibitem[Darve(2007)]{darve2007therm}
Darve,~E. \emph{Free Energy Calculations}; Springer, 2007; pp 119--170\relax
\mciteBstWouldAddEndPuncttrue
\mciteSetBstMidEndSepPunct{\mcitedefaultmidpunct}
{\mcitedefaultendpunct}{\mcitedefaultseppunct}\relax
\EndOfBibitem
\end{mcitethebibliography}

\end{document}